\documentclass{aa} 
\usepackage{graphicx}
\usepackage{txfonts}
\begin{document} 

   \title{Ca II Triplet Spectroscopy of Small Magellanic Cloud Red Giants. VI. Analysis of chemical properties of the Main Body}
    \titlerunning{CaT spectroscopy of SMC red giant stars}
    
   \author{B.J. De Bortoli\inst{1,2,3}
          \and
          M.C. Parisi\inst{3,4,5}
          \and 
          L.P. Bassino\inst{1,,2,3}
          \and 
          D. Geisler\inst{6,7,8}
          \and 
          B. Dias\inst{9}
          \and 
          G. Gimeno\inst{10}
          \and
          M.S. Angelo\inst{11}
          \and
          F. Mauro\inst{12}
          }

\institute{
        Facultad de Cs. Astronómicas y Geofísicas, Universidad Nacional de La Plata, Paseo del Bosque s/n, 1900, La Plata, Argentina.\\
            \email{brudebo.444@gmail.com}
            \and
            Instituto de Astrofísica de La Plata (CCT La Plata, UNLP-CONICET), Paseo del Bosque s/n, 1900, La Plata, Argentina.
         \and
            Consejo Nacional de Investigaciones Cient\'ificas y T\'ecnicas, Godoy Cruz 2290, C1425FQB,  
Ciudad Aut\'onoma de Buenos Aires, Argentina.
         \and   
         Observatorio Astron\'omico de C\'ordoba, Universidad Nacional de Córdoba, Laprida 854, X5000BGR, Córdoba, Argentina.
         \and
           Instituto de Astronomía Teórica y Experimental (UNC-CONICET), Laprida 854, X5000BGR, Córdoba, Argentina.
         \and
        Departamento de Astronomia, Casilla 160-C, Universidad de Concepcion, 4030000, Concepción, Chile.
         \and
        Instituto de Investigación Multidisciplinario en Ciencia y Tecnología, Universidad de La Serena. Avenida Raúl Bitrán S/N, La Serena, Chile.
         \and
        Departamento de Astronomía, Facultad de Ciencias, Universidad de La Serena. Av. Juan Cisternas 1200, 1720236, La Serena, Chile.
          \and
          Instituto de Alta Investigación, Sede Esmeralda, Universidad de Tarapacá, Av. Luis Emilio Recabarren 2477, 1100000, Iquique, Chile.
          \and
       Gemini Observatory, NSF’s NOIRLab, 950 N. Cherry Ave., Tucson, AZ 85719, USA
          \and
          Centro Federal de Educação Tecnológica de Minas Gerais, 38778-000, Minas Gerais, Brasil.
          \and
       Instituto de Astronomía, Universidad Católica del Norte, 127 0236, Antofagasta, Chile.
   }
 
   \date{Received ...; accepted ...}
  

 \abstract%
{}
   {We aim to analyze the chemical evolution of the Main Body of the SMC, adding six additional clusters to previously published samples, based on homogeneously determined and accurate metallicities.}
   {We derived radial velocities and Ca II Triplet (CaT) metallicity of more than 150 red giants stars in six SMC star clusters and their surrounding fields, with the instrument GMOS on GEMINI-S. The mean cluster radial velocity and metallicity were obtained with mean errors of 2.2 km\,s$^{-1}$ and 0.03 dex, while the mean field metallicities have a mean error of 0.13 dex. We add this information to that available for another 51 clusters and 30 fields with CaT metallicities on the same scale. Using this expanded sample we analize the chemical properties of the SMC Main Body, defined as the inner 3.4 degrees in semimajor axis.}
   {We found a high probability that the metallicity distribution of the Main Body clusters is bimodal with a  metal-rich and a metal-poor cluster group, having mean metallicities with a dispersion of $\mu = -0.80$, $\sigma = 0.06$ and $\mu = -1.15$, $\sigma = 0.10$ dex, respectively.  On the other hand, Main Body field stars show a unimodal metallicity distribution peaking at $[Fe/H] \sim -1$ and dispersion of $0.3$. Neither metal-rich nor metal-poor clusters present a metallicity gradient. However the full Main Body cluster sample and field stars have a negative metallicity gradient consistent with each other, but the one corresponding to clusters has a large error due to the large metallicity dispersion present in the clusters studied in that region. Metal-rich clusters present a clear age-metallicity relation, while metal-poor clusters present no chemical enrichment throughout the life of the galaxy.}
   { We present observational evidence that the chemical enrichment is complex in the SMC Main Body. Two cluster groups with potential different origins could be coexisting in the Main  Body.
More data with precise and homogeneous metallicities and distances are needed and dynamical simulations are required to understand possible different origins for the two possible cluster groups.
}

   \keywords{Galaxies: star clusters: general  --
                Magellanic Clouds  --
                stars: abundances 
               }

\maketitle
\section{INTRODUCTION}
\label{sec.intro}
The Magellanic System \citep{donghia+16} is one of the most rewarding nearby systems for the study of dwarf galaxies and the stellar populations that they host. It consists of the Small Magellanic Cloud (SMC), the Large Magellanic Cloud (LMC), the Bridge, the Magellanic Stream and the Leading Arm. The SMC and LMC are the closest pair of interacting galaxies to the Milky Way (MW) located at distances of 49.59 $\pm$ 0.09 kpc \citep{pietrzynski+19} and 62.44 $\pm$ 0.47 kpc \citep{Graczyk+20}, respectively. The Magellanic Clouds (MCs) are embedded within a diffuse structure of HI gas \citep{mathewson+74,putman+03,nidever+08,nidever+10} which has been interpreted by numerous works as the consequence of interaction either between SMC and LMC or among SMC, LMC and the MW \citep[e.g.,][]{diaz+11,diaz+12,besla+12,besla+16}. There is some discussion in the literature as to whether MCs have been orbiting the MW \citep{gardiner+96,diaz+12}, however the latest research based on the most accurate measurements of proper motions with the
Hubble Space Telescope (HST) and Gaia, and updated LMC and Milky Way mass estimates, suggests that they are experiencing their first encounter with our Galaxy \citep{besla+07,besla+10,piatek+08,kallivayalil+13,patel+17,gaia+18}.

Regardless of whether the MCs are experiencing their first encounter with the MW or not, it is expected that the morphology of both galaxies is being modified by tidal interactions \citep{martin+21}. The dynamical simulations carried out in recent years, e.g. \citet{connors+06,bekki+07,diaz+12,besla+12}, have been able to reproduce many of the morphological characteristics of the Magellanic System (Magellanic Stream, Leading Arm, Magellanic Bridge, Counter-Bridge) as a consequence of the interaction between the MCs. Large numbers of studies have found evidence of tidal tails around the SMC and the LMC \citep{besla+07,belokurov+19,nidever+19,gaia+21,ElYoussoufi+2021,dias+21}. In particular,  the complex patterns of velocities in the SMC found, for example, by \citet{niederhofer+18} and \citet{Niederhofer+21}, suggest that this galaxy is being tidally disrupted by the LMC. Also, the current morphology of the SMC could be the consequence of a recent collision between the MCs \citep{besla+12,zivick+18}. 

Given such interactions, the star formation histories and chemical enrichment processes of the stellar populations of a galaxy are of course affected \citep{whitmore+99,dacosta91,dopita+97,pagel+98}. Their spatial, age and metallicity distributions and gradients present distinctive effects of the interaction processes \citep[e.g.,][]{Cioni09,dobbie+14a,nayak+16,dias+16b,rubele+18,deleo+20,santos+20}. SMC star clusters have proven historically to be excellent tracers of the chemical and dynamical history of this galaxy \citep[e.g.,][]{dacosta+98,glatt+10,parisi+09,parisi+14,parisi+15,perren+17,piatti18,nayak+18,bitsakis+18,narloch+21,dias+21}. However, despite the exhaustive study of the chemical properties of the SMC clusters, there is some controversy in the literature regarding the chemical evolution of this galaxy using those objects as tracers.

Several studies have shown that the field metallicity distribution (MD)  is unimodal with a [Fe/H] maximum near -1 dex \citep[e.g.,][]{carrera+08,parisi+10,dobbie+14b,parisi+16,choudhury+18,choudhury+20}. However, using a sample of 36 clusters homogeneously studied, \citet{parisi+15} (hereafter P15) suggest that the MD could be bimodal with a probability of 86\%. This probability drops drastically to 59\% when the cluster sample increases significantly \citep[][hereafter P22]{parisi+22}, but nevertheless the sample shows a marked absence of clusters in the internal region with metallicity values typical of the SMC field. 

Although there is agreement between most studies regarding the existence of a metallicity gradient (MG) in the SMC field \citep[e.g.,][]{carrera+08,parisi+10,dobbie+14b,parisi+16,choudhury+18,choudhury+20} it is not clear if the SMC clusters present such a gradient. \citet{narloch+21} studied 35 clusters with Strömgren photometry and they found that younger, more metal-rich star clusters are concentrated mainly towards the centre of the galaxy, while older, more metal-poor clusters are located further from the centre. However, studies based on CaII triplet (CaT) metallicities from \citet{parisi+09} (hereafter P09), P15, \citet{dias+22} (hereafter D22) and P22 find that although there is a tendency for the clusters to be more metal-poor as we move away from the centre of the galaxy up to 4$^{\circ}$, the cluster MG is not statistically significant due to the large cluster metallicity dispersion in the internal region ($\sim$ 0.6 dex, P15, P22). To make things more interesting, employing the separation of SMC cluster samples into the different sky regions defined by \citet[][hereafter D16,D21]{dias+16b,dias+21}, we can see that the clusters belonging to the Northern Bridge appear to be the ones that best trace a V-shape in the MD (P22, D22) and the West Halo clusters could present a MG (D16, D22) but with some uncertainty (P22).  

The mentioned metallicity dispersion, not only observed in spectroscopic studies but also in photometric ones \citep[e.g.,][]{perren+17,narloch+21}, is also evident in the analysed age-metallicity relationships (AMR)  in which the models of chemical evolution proposed in the literature for the SMC do not reproduce the data in general \citep{dacosta+98,harris+04,pagel+98,carrera+08,cignoni+13,tsujimoto+09,perren+17}. Also, the V-shape present in the field and metallicity gradient \citep[P15,P16,][D16]{bica+20} is under discussion, specially in the outer region, where it is no clear that the gradient increases or remains constant \citep[][P22, D22]{parisi+16,choudhury+20}. 

 Galaxy-galaxy interactions are expected to mostly affect the outskirts of the SMC \citep{mayer+01}, but D22 show that the SMC disruption due to tidal effects starts much further inside the SMC tidal radius \citep{dias+21}. P22 also showed that all the components suggested by D21 present a minimum in metallicity as well as smaller metallicity dispersion near the projected tidal radius. Significantly different cluster chemical properties are displayed by samples inside vs. outside the tidal radius. The outer cluster system of the SMC is being systematically studied by the VISCACHA survey \citep{maia+19} with results that impose important constraints on dynamical models of the Magellanic Clouds (D21, D22). In this paper we focus our analysis in the internal region, where chemical evolution does not seem to follow a canonical behavior either. 

We organize the paper as follows: In Section \ref{sec.obs} we describe the cluster sample selection, observations and data reduction process. The measurement of radial velocities and equivalent width of the Ca II Triplet lines and the metallicity determination are presented in Section \ref{sec.parameters}. Sections \ref{sec.memb} and \ref{sec.met.cl} are dedicated to the membership and metallicity analysis, respectively. Finally we summarize our results in Section \ref{sec.conclu}.

\section{OBSERVATIONS AND DATA REDUCTION}
\label{sec.obs}

\begin{figure}
  \centering
\includegraphics[width=85mm]{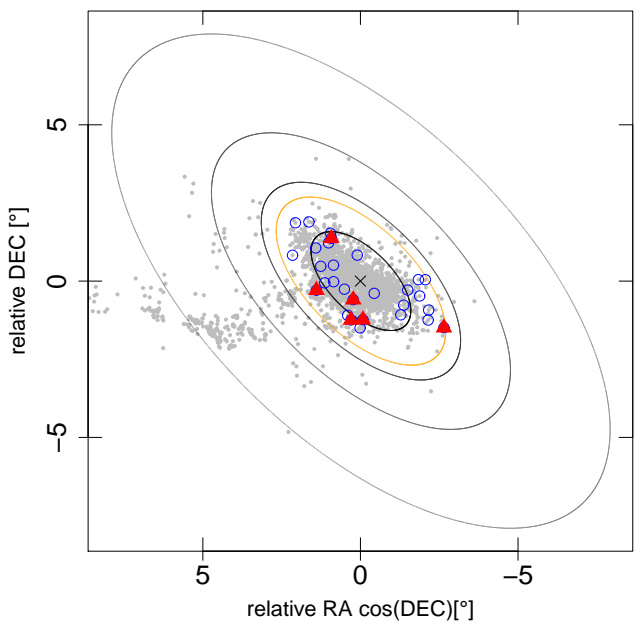}
\caption{Projected spatial distribution of SMC star clusters. Gray dots show those catalogued by \citet{bica+20}. Blue open circles are star clusters taken from the literature (\citet{dacosta+98}, P09, P15, D21, P22 and D22)  and filled red triangles are those studied in this paper. The orange ellipse corresponds to a semi-major axis of $3.4^{\circ}$. Black ellipses correspond to semi-major axes of $2^{\circ}$, $4^{\circ}$, $6^{\circ}$ and $10^{\circ}$. They are centred on the SMC, with PA=$45^{\circ}$ and b/a=1/2.}
\label{fig.espa}
\end{figure}

In order to increase the number of inner SMC star clusters homogeneously studied with the CaT technique, we have observed 6 clusters in the inner region of the SMC, which are spatially distributed as can be seen in Figure\,\ref{fig.espa}. The ellipses are defined in the plane of relative RA vs. relative DEC, with ratio b/a=0.5 and position angle = 45$^{\circ}$ \citep{piatti+05,dias+14}. The semi-major axis $a$ of the ellipse coincident with the position of the cluster is used as the projected distance from the SMC center. We considered the semi-major axis $a =$ 3.4$^{\circ}$ as the division between the inner and outer regions of the SMC (D21), so our sample basically consists of clusters in the SMC Main Body (D16, D21, P22). 
Clusters were selected from \citet{piatti11a}, \citet{piatti11b} and \citet{piatti+11}.  We have prioritized those clusters whose Color-Magnitude Diagram (CMD) show the most populated red giant branch (RGB), in order to maximize the number of suitable targets for the CaT technique.  
Although our whole sample has ages and metallicities previously determined from photometric techniques, in this work we provide more accurate spectroscopic metallicities derived from the CaT technique as well as radial velocities (RVs). The cluster sample is presented in Table\,\ref{tab.oursample}, which includes coordinates, projected distances and ages. 

\begin{table*}[!htb]
\caption{Selected SMC cluster sample.}             
\label{tab.oursample}      
\centering                          
\footnotesize
\begin{tabular}{l c c c c }        
\hline\hline                 
\noalign{\smallskip}
  \multicolumn{1}{l}{Cluster} &
  \multicolumn{1}{c}{RA (J2000.0)} &
  \multicolumn{1}{c}{DEC (J2000.0)} &
  \multicolumn{1}{c}{a} &
  \multicolumn{1}{c}{Age} \\
    \multicolumn{1}{c}{} &
  \multicolumn{1}{c}{($h$ $m$ $s$)} &
  \multicolumn{1}{c}{($^{\circ}$ ' '')} &
  \multicolumn{1}{c}{($^{\circ}$)} &
  \multicolumn{1}{c}{(Gyr)} \\
\noalign{\smallskip}
\hline\hline
\noalign{\smallskip}
K\,38, L\,57 & 0:57:49.5 & -73:25:23 & 1.363 & $3.0\pm0.4$$^1$\\
HW\,31, [RZ2005]\,97 & 0:55:34.0 & -74:3:46 & 2.149 & $4.6\pm0.3$$^1$\\
HW\,41, [RZ2005]\,125 & 1:00:33.6 & -71:27:13 & 1.769 & $5.6\pm0.4$$^1$\\
HW\,42  & 1:01:08.0 & -74:04:25 & 2.617 & $\approx2.4$$^2$\\
L\,3, ESO\,28-13, OGLE-CL\,SMC\,323 & 0:18:25.2 & -74:19:05 & 2.938 & $1.2\pm0.3$$^3$\\
L\,91, [RZ2005]\,194  & 1:12:51.6 & -73:07:07 & 2.609 & $4.1\pm0.3$$^1$\\
\noalign{\smallskip}
\hline                                   
\end{tabular}\\
\textbf{References:} $^1$\citet{parisi+14}, $^2$Bica et al. (2022, in prep.),$^3$\citet{dias+14}
\end{table*}

\begin{figure}
  \centering
\includegraphics[width=70mm]{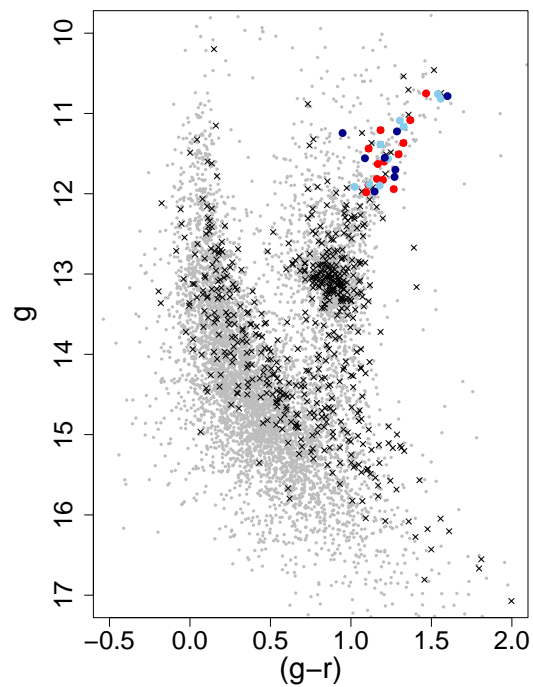}
\caption{Color- magnitude diagram of the cluster K\,38. Gray small circles show all the objects located in the corresponding frame. Black crosses are those located inside the cluster radius. Blue, cyan and green symbols represent targets discarded due to their distance from the cluster centre, RV and metallicity, respectively. Red circles are cluster members according to our membership analysis (see Section \ref{sec.memb} for details).} 
\label{fig.cmd}
\end{figure}

For each cluster, pre-images in the $g$ and $r$ filters had been previously obtained with the instrument Gemini/GMOS-S (programme GS-2014B-Q-78), from which the CMDs [$g$, $g-r$] were built. We selected RGB stars from the CMDs as spectroscopic targets. As an example, we show in Figure\,\ref{fig.cmd} the CMD for the cluster K\,38. We marked the spectroscopic targets with large circles in the figure, following the color code related to our membership analysis (see the figure caption and Section \ref{sec.memb} for details).

The spectroscopic data for the selected RGB stars consist of intermediate resolution infrared spectra obtained with GMOS-S \citep{hook+04}, with the new Hamamatsu CCDs featuring enhanced red sensitivity \citep{gimeno+16} in MOS mode (programme GS-2016B-Q-17). It comprises more than 150 spectra of stars in the area of our SMC cluster sample.  Observations of 4$\times$900$\,{\rm sec}$ exposure time and 2$\times$2 binning were taken for each frame, using the R831 disperser and the CaT\_G0333 filter. The four observations were taken in pairs centred on $8500$\,\AA\,and $8550$\,\AA\,for dithering. In average, 26 slits of 1$\,{\rm arcsec}$ width were located in each frame. This instrumental configuration yields a spectral resolution of 0.075\,${\rm nm/px}$ (R $\sim$ 2000). Calibration observations, such as bias, flat fields and CuAr arc spectra, were also taken.

In order to reduce the spectra, we used the script\footnote{http://drforum.gemini.edu/topic/gmos-mos-guidelines-part-1/} developed by M. Angelo. First, we applied the bias and flat corrections, interpolated across the useless CCDs gap (\textmd{GMOSAIC} task), identified the spatial extension of each slitlet (\textmd{GSCUT} task), and cut and pasted the two-dimensional spectra in different FITS file extensions. Additionally, bad pixel masks have been applied. The wavelength solution for each individual spectrum has been obtained from \textmd{GSWAVELENGTH} task and differences in quantum efficiency between the 3 CCDs chips have been corrected with the \textmd{GQECORR} task. Afterwards, we used \textmd{CRMEDIAN} and \textmd{FIXPIX} to remove cosmic rays. We then ran the  \textmd{GSTRANSFORM} task, using arc spectra, to rectify the spectra and apply the appropriate wavelength solution. Finally, we extracted and combined the spectra of different central wavelengths, summing them. 

In order to analyse if a zero point correction to our wavelength calibration was necessary, we measured in our spectra the centre of several bright sky lines with well known central wavelengths \citep{hanuschik03}. When a non-negligible difference was found, we used \textmd{SPECSHIFT} to shift the spectra. In all cases corrections were smaller than 0.4\,\AA\,($\sim 14$ ${\rm km\,s^{-1}}$), except for the cluster K\,38 for which it was necessary to apply a shift of 0.9\,\AA \,($\sim 32$ ${\rm km\,s^{-1}}$).

\begin{figure}
  \centering
\includegraphics[width=85mm]{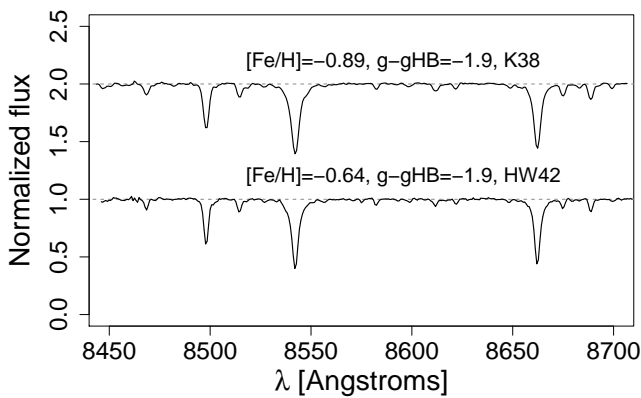}
\caption{Normalized GMOS spectra for the clusters K\,38 (top) and HW\,42 (bottom).}
\label{fig.espectro}
\end{figure}

Finally, we ran the \textmd{CONTINUUM} task to normalize the spectra to the pseudo-continuum level \citep[e.g.,][]{vasquez+15}. As an example, we show in Figure\,\ref{fig.espectro} two normalized spectra of RGB stars in two clusters with very similar evolutionary stages but different metallicities. The signal-to-noise ratio (SNR) for combined and normalized spectra range from 50 to 130.

\section{STELLAR PARAMETERS DETERMINATION}
\label{sec.parameters}
\subsection{Radial velocity and equivalent width measurements}
\label{sec.rv}
The determination of target radial velocities (RVs) is crucial for cluster membership determination. They are also needed to perform the Doppler correction in order to measure equivalent widths (EW) of the three CaT lines in rest spectra.

In order to measure RVs of our program stars, we cross-correlated our normalized spectra with a RGB theoretical template taken from the library of \citet{coelho14}. We selected a template with typical parameters for a RGB star ($T_{eff}$=5000$\,{\rm K}$, $log(g)$=1$\,{\rm dex}$, $[Fe/H]$=-1$\,{\rm dex}$ and $[\alpha/Fe]$=0.4$\,{\rm dex}$), and we convolved the spectrum to reach the spectral resolution of GMOS. As was shown by D22, the mean RV does not vary significantly when adopting different templates in the ranges of Teff = 4700 to 5300 K, [Fe/H] = -1.3 to -0.5, and log(g) = 1, which are typical of the RGB stars.

We used the \textsc{IRAF} task \textmd{FXCOR} to perform the cross-correlation, which additionally corrects the observed RVs providing the heliocentric values. We obtained a typical RV error of $\sim$\,4\,${\rm km\,s^{-1}}$.

As we describe in the next section, EWs of the CaT lines are required for the metallicity determination. Following the same procedure as in our previous work (e.g., P09, P15), the pseudo-continuum was fitted in a pair of continuum windows to shorter and longer wavelengths of the corresponding line centre, using the line and continuum bandpasses from \citet{armandroff+88}. We fitted a Gaussian plus a Lorentzian function to each CaT line, with respect to the pseudo-continuum, and calculated the "pseudo-EW". As shown by \citet{rutledge+97a,rutledge+97b} and \citet{cole+04}, this combined function takes into account adequately the contribution of the wings and the core of the line profile. We obtained a typical EW error of\,$\sim$0.08 {\AA}.

\subsection{Metalicity determination}
\label{sec.met}
Many studies have calibrated the strength of the CaT lines with metallicity from integrated and individual spectra of RGB stars. In the case of individual stellar spectra, the technique requires the construction of the so-called CaT index as the sum of the EW of two or three CaT lines ($\Sigma$EW). Also, it is necessary to remove the effects of surface gravity and temperature on the $\Sigma$EW \citep{armandroff+91,olszewski+91}, e.g. by using the difference in magnitudes, in a given filter, between the observed star and the horizontal branch or red clump. A detailed description of the different ways of building the CaT index and the filters used in the literature, as well as the available calibrations, can be found in \citet[][hereafter DP20]{dias+20} and references therein.

Usually, using the sum of the three CaT lines is the best choice because it takes into account the complete information. However, if the SNR is low, the faintest line adds mostly noise and increases the errors in metallicity, so that it is avoided in these cases, following an adequate calibration. In this work, most of the spectra have SNR between 50 and 130, which is high enough to produce good quality CaT lines in almost all cases. Therefore we use the calibration including the three lines, as follows:

\begin{equation}
\Sigma EW = EW_{8498} + EW_{8542} + EW_{8662}
\end{equation}

 In those cases in which the weakest line could not be well fit, we add the contribution of the two most intense lines and then performed the corresponding conversion according to equation 5 of DP20. 
 We then calculated the reduced EW ($W'$) from: 

\begin{equation}
W' = \Sigma EW + \beta (g - g_{HB})
\end{equation}

\noindent where $(g - g_{HB})$ represents the difference between the magnitude of the star ($g$) and the magnitude of the cluster's horizontal branch/red clump ($g_{HB}$). By considering the magnitude difference, we not only remove the dependence of $\Sigma$EW with surface gravity and temperature, but also its dependence on cluster distance and interstellar reddening. We adopt a  $\beta$ value of 0.85$\pm$0.08 from DP20, which was calculated for filter g using as reference $\beta_V = 0.71$ for the canonical V filter. We note that this value of $\beta_V$ is the same used in our previous works, therefore our measurements are on the same scale.\\

\begin{figure}
  \centering
\includegraphics[width=85mm]{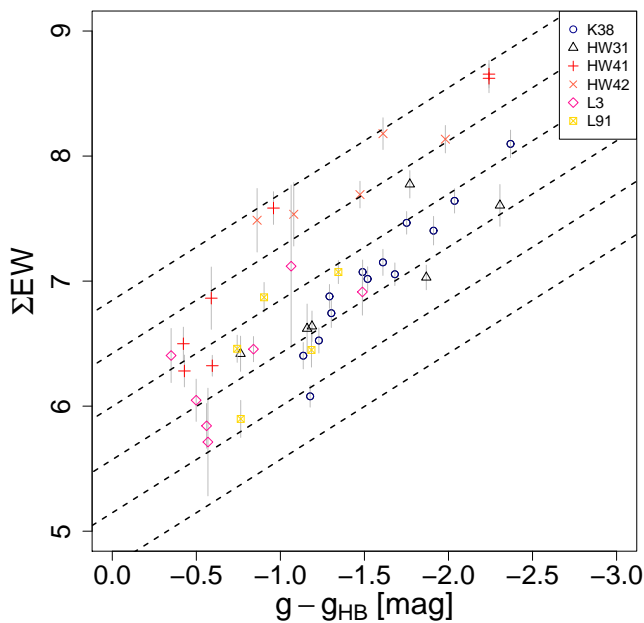}
\caption{Sum of the equivalent width of the three CaT lines against $g$-$g_{HB}$ for member stars of the six SMC clusters. Dashed lines represent lines of constant metallicity corresponding to $[Fe/H]$= $-0.5$, $-0.65$, $-0.80$, $-0.95$, $-1.10$ and $-1.25$ ${\rm dex}$, from top to bottom.}
\label{fig.sumdif}
\end{figure}

The $g$ magnitudes were obtained from PSF photometry performed on the pre-images using the  SkZ pipeline \citep{mauro+13,mauro20}. We then determined $g_{HB}$ from the cluster CMD following the procedure detailed in DP20. Figure\,\ref{fig.sumdif} shows the $\Sigma$EW as a function of $g$-$g_{HB}$ for targets that resulted to be cluster members according to our membership analysis (see Section \ref{sec.memb}).

Finally, we calculated the metallicity of each observed star using the calibration derived by DP20,

\begin{equation}
[Fe/H] = -2.917 (\pm 0.116) + 0.353 (\pm 0.020) W'
\end{equation}

As shown by DP20, their calibration is in excellent agreement with the one derived by \citet{cole+04}. Therefore, our metallicity  values are on the same scale as our previous work, in which Cole's calibration was used. We have obtained individual uncertainties in [Fe/H] of typically\,$\sim$0.2 ${\rm dex}$.\\

\begin{table*}[!htb]
\caption{Measured parameters for the observed RGB stars}             
\label{tab.parameters}      
\centering                          
\footnotesize
\begin{tabular}{l c c c c c c c}        
\hline\hline                 
\noalign{\smallskip}
  \multicolumn{1}{l}{Star ID} &
  \multicolumn{1}{c}{RA (J2000.0)} &
  \multicolumn{1}{c}{DEC (J2000.0)} &
  \multicolumn{1}{c}{RV} &
  \multicolumn{1}{c}{g$-g_{HB}$} &
\multicolumn{1}{c}{$\Sigma$EW } &
\multicolumn{1}{c}{$[Fe/H]$} &
\multicolumn{1}{c}{Cluster/Field} \\
    \multicolumn{1}{c}{} &
  \multicolumn{1}{c}{($h$ $m$ $s$)} &
  \multicolumn{1}{c}{($^{\circ}$ ' '')} &
  \multicolumn{1}{c}{${\rm (km\,s^{-1})}$} &
\multicolumn{1}{c}{${\rm (mag)}$} &
\multicolumn{1}{c}{({\AA})} &
\multicolumn{1}{c}{${\rm (dex)}$} &
    \multicolumn{1}{c}{C/F} \\
\noalign{\smallskip}
\hline\hline
\noalign{\smallskip}
K\,38-2 & 0:57:49.3 & -73:22:45 & $177.2\pm4.2$ & -1.56 & $7.88\pm0.13$ & $-0.60\pm0.21$ & F \\
K\,38-4 & 0:58:15.9 & -73:26:09 & $161.7\pm2.8$ & -1.57 & $7.27\pm0.09$ & $-0.82\pm0.20$ & F \\
\noalign{\smallskip}
\hline                                   
\end{tabular}\\
\textbf{Note:} Full table is available online.
\end{table*}

Table\,\ref{tab.parameters} presents coordinates and measured values, with their respective errors, for all observed RGB stars. We have differentiated the cluster member stars from those belonging to their surrounding fields, according to our membership analysis (see section\,\ref{sec.memb}). Stars are labelled with the cluster name plus a number, which represents the aperture number in our program.

\section{MEMBERSHIP ANALYSIS}
\label{sec.memb}
\begin{figure}
  \centering
\includegraphics[width=80mm]{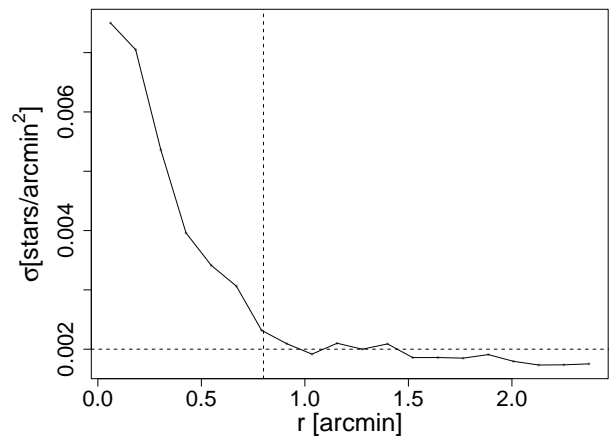}
\caption{Radial stellar density profile of the cluster L91. Horizontal dashed line  is the stellar background level and the vertical dashed line shows the adopted cluster radius. 
}
\label{fig.radial_profile}
\end{figure}

\begin{figure*}
  \centering
\includegraphics[width=140mm]{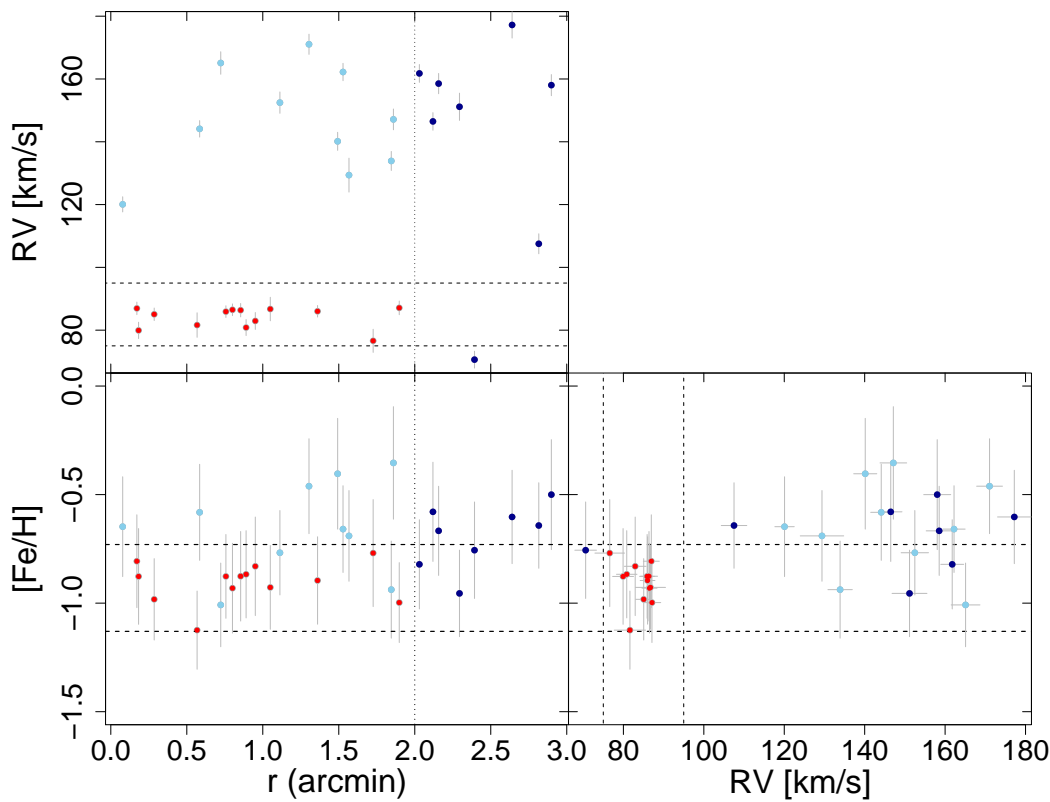}
\caption{Membership analysis for the cluster K\,38. The color code is the same as in Figure\,\ref{fig.cmd}. Cluster radius is shown by the dotted vertical line and the dashed lines represent the adopted RV and metallicity cuts. }
\label{fig.params}
\end{figure*}

In order to separate cluster members from SMC field stars, we followed the same procedure as in our previous work (see P09, P15 and P22 for more details). In summary, for each cluster, we first built the radial density profile using stellar counts from our photometry and adopting the cluster centre from \citet{bica+20}. The cluster radius is defined as the distance from the centre out to the level where the profile intersects the background density, but in some cases we adopted a smaller radius in order to maximize the probability of the selected stars to belong to the clusters  (Figure \ref{fig.radial_profile}). We assume the background level to be that at which the radial profile does not decease significantly any further. The exception are the clusters HW\,42 and K\,38 for which the radial profile could not be constructed due to the low cluster overdensity with respect to the field. In these cases, the radius values from \citet{bica+20} were adopted. We then analysed the targets RV and metallicity as a function of the distance from the cluster centre (Figure\,\ref{fig.params}). Stars closer to the centre have a larger probability of being cluster members. It is also assumed that, in general, cluster members have smaller velocity dispersion and potentially different mean  RV and metallicity compared to field stars. We adopted the RV and metallicity cuts of $\pm 10 {\rm km\,s^{-1}}$ and $\pm 0.2 {\rm dex}$ (from C09, P15 and P22) around the mean value of the visually identified probable members candidates. The RV cuts are consistent with the intrinsic cluster RV dispersion and our mean RV error. The adopted metallicity cuts are representative of our mean metallicity error.\\

We consider as member stars those targets located closer to the cluster centre than the adopted radius and having RV and metallicity values inside the cuts (red symbols in Figure\,\ref{fig.params}). In the figure, targets discarded as probable cluster members because of their distance, RV and metallicity values are represented with blue, cyan and green circles, respectively. The RV and metallicity cuts are shown with short dashed lines in Figure\,\ref{fig.params} while the adopted cluster radius is represented by the dotted line.\\

Finally, we checked the proper motions (PMs) of our targets, from the Gaia EDR3 \citep{gaia+21} catalogue, in order to maximize the membership probability of our cluster member candidates. We use the PM to verify that the spectroscopic members have consistent PM values among them. None of the stars considered cluster members, according to the criteria described above, was discarded due to their PMs. \\

Using only member stars we calculated the simple mean RV and metallicity for each cluster. As a representative metallicity of the cluster surrounding fields, we calculated the median of nonmember stars metallicity values, as suggested by \citet{dobbie+14b,parisi+16}. To determine the median field metallicities, we verified that the selected stars do not have RV and metallicity values close to the limits imposed by the adopted cuts nor were located close to the cluster radius ($\Delta r < 0.2$ $arcmin$, $\Delta RV < 5$ $km\,s^{-1}$ and $\Delta [Fe/H] < 0.1$ $dex$). The results for both cluster and field stars are given in Table\,\ref{tab.final}. The median metallicity for the field of L\,3 is less reliable because the sample contains only three stars.

\section{CHEMICAL PROPERTIES OF CLUSTER AND FIELD STARS}
\label{sec.met.cl}

\begin{table*}[!htb]
\caption{Mean parameters for the selected SMC star clusters and their standard error of the mean.}             
\label{tab.final}      
\centering                          
\footnotesize
\begin{tabular}{l c c c c c}        
\hline\hline                 
\noalign{\smallskip}
  \multicolumn{1}{l}{Cluster} &
 \multicolumn{1}{c}{n} &
  \multicolumn{1}{c}{RV} &
  \multicolumn{1}{c}{[Fe/H]} &
  \multicolumn{1}{c}{n$_{\rm field}$} &
  \multicolumn{1}{c}{[Fe/H]$_{\rm field}$ ($\sigma$)} \\
    \multicolumn{1}{c}{} &
    \multicolumn{1}{c}{} &
  \multicolumn{1}{c}{${\rm (km\,s^{-1})}$} &
  \multicolumn{1}{c}{${\rm (dex)}$} &
  \multicolumn{1}{c}{} &
\multicolumn{1}{c}{${\rm (dex)}$} \\
\noalign{\smallskip}
\hline\hline
\noalign{\smallskip}
K\,38 & 13 & $84.0\pm0.9$ & $-0.90\pm0.02$ & 18 & $-0.65\pm0.04$ (0.18) \\
HW\,31 & 6 & $125.5\pm3.4$ & $-0.89\pm0.04$ & 15 & $-1.12\pm0.10$ (0.37) \\
HW\,41 & 7 & $143.6\pm1.6$ & $-0.67\pm0.05$ & 14 & $-0.96\pm0.10$ (0.36) \\
HW\,42  & 5 & $144.3\pm2.0$ & $-0.58\pm0.03$ & 13 & $-0.95\pm0.12$ (0.42) \\
L\,3 & 7 & $140.1\pm3.4$ & $-0.90\pm0.05$ & 3 & $-0.75\pm0.37$ (0.33) \\
L\,91 & 5 & $126.7\pm1.8$ & $-0.90\pm0.06$ & 18 & $-1.01\pm0.08$ (0.35) \\
\noalign{\smallskip}
\hline                                   
\end{tabular}\\
\end{table*}

In order to perform a statistically more significant study of the SMC chemical history, we enlarged our cluster sample by adding 51 clusters having CaT metallicities derived with the same method as in this work: 1 cluster from  \citet[][hereafter DH98]{dacosta+98}, 15  from P09, 13 from P15, 7 from D21, 12 from P22 and 3 from D22. Out of the 7 clusters studied by DH98 we only included here NGC\,121. For the rest of the DH98 cluster sample we adopt, for homogeneity, the values derived by P15 and P22, which are in excellent agreement with the CaT metallicities derived by DH98. P09, P15 and P22 employed data from the VLT-FORS2 whereas clusters studied in P22 and in this work were observed with Gemini-GMOS. The FORS2 and GMOS samples have two clusters in common (NGC\,151 and K\,8). The FORS2 metallicity of NGC\,151 (P22) is in very good agreement with the GMOS metallicity (D22). In the case of K\,8 both datasets (P15 and D22) have four stars in common with metallicities in agreement, one of them a cluster member. Therefore, we consider that the metallicities based on FORS2 and GMOS data are consistent. Then, our final sample includes 57 clusters with metallicities on the same scale spread over all SMC regions of which 37 are in the Main Body. Hence, the 6 GCs analysed in this work represent an increase in the inner SMC cluster sample by 16\%. \\ 

In addition, we add to our field sample those studied by \citet[][]{parisi+10} and \citet[][]{parisi+16}, hereafter P10 and P16. The fields studied in these two works (15 from P10 and 15 from P16) correspond to  the fields surrounding the clusters studied in P09 and P15, respectively. Therefore we have a total of 36 SMC fields with homogeneously determined average metallicities.\\

\subsection{Metallicity Distribution}
\label{subsec.MD}

\begin{figure}
  \centering
\includegraphics[width=85mm]{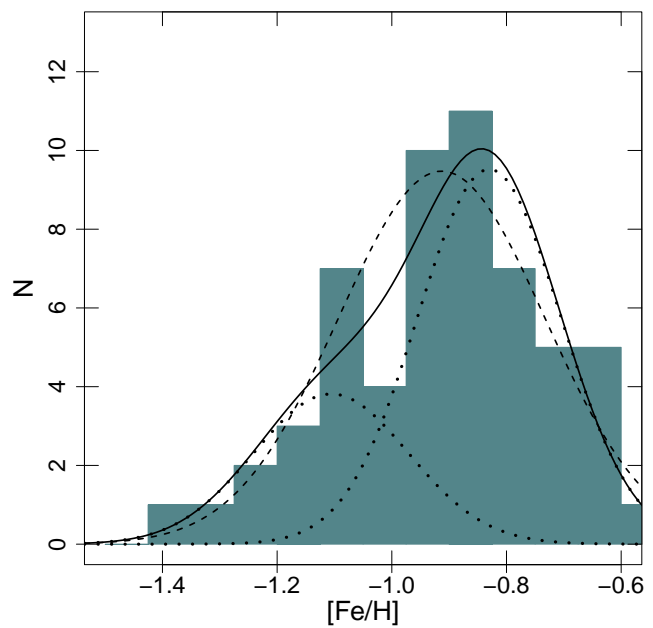}
\caption{Metallicity distribution of the enlarged cluster sample. Dotted lines are the two fitted Gaussian functions according to the results from the GMM test and the solid line shows their sum. Dashed line shows the fitted Gaussian in the unimodal case.}
\label{fig.MD}
\end{figure}

In order to analyse the existence of bimodality in the MD, as  suggested by P09 and P15, we applied the Gaussian Mixture Modeling test \citep[GMM][]{muratov+10} to the full sample. The MD of our cluster sample is shown in Figure\,\ref{fig.MD}, as well as the fits considering one (dashed line) or two (dotted lines) Gaussian functions. The results for the application of this algorithm  are summarized in the first line of Table \ref{tab.gmm}. In the table we list the sample, the peaks ($\mu$) and the $\sigma$ for the unimodal and bimodal fits, the p value (which is the probability of being wrong in rejecting unimodality), the bimodality probability given by the parametric bootstrap, the separation of the peaks of the fitted Gaussian functions (DD) and the kurtosis of the distribution (k). An indication of a bimodal distribution is given by a negative k value and $DD > 2$ \citep{ashman+94}. As can be seen from the table we found a significantly lower probability (39\%) than that obtained by P15 (86\%) that the whole SMC cluster MD is bimodal, in agreement with P22 (59\%). \\

\begin{figure}
  \centering
\includegraphics[width=85mm]{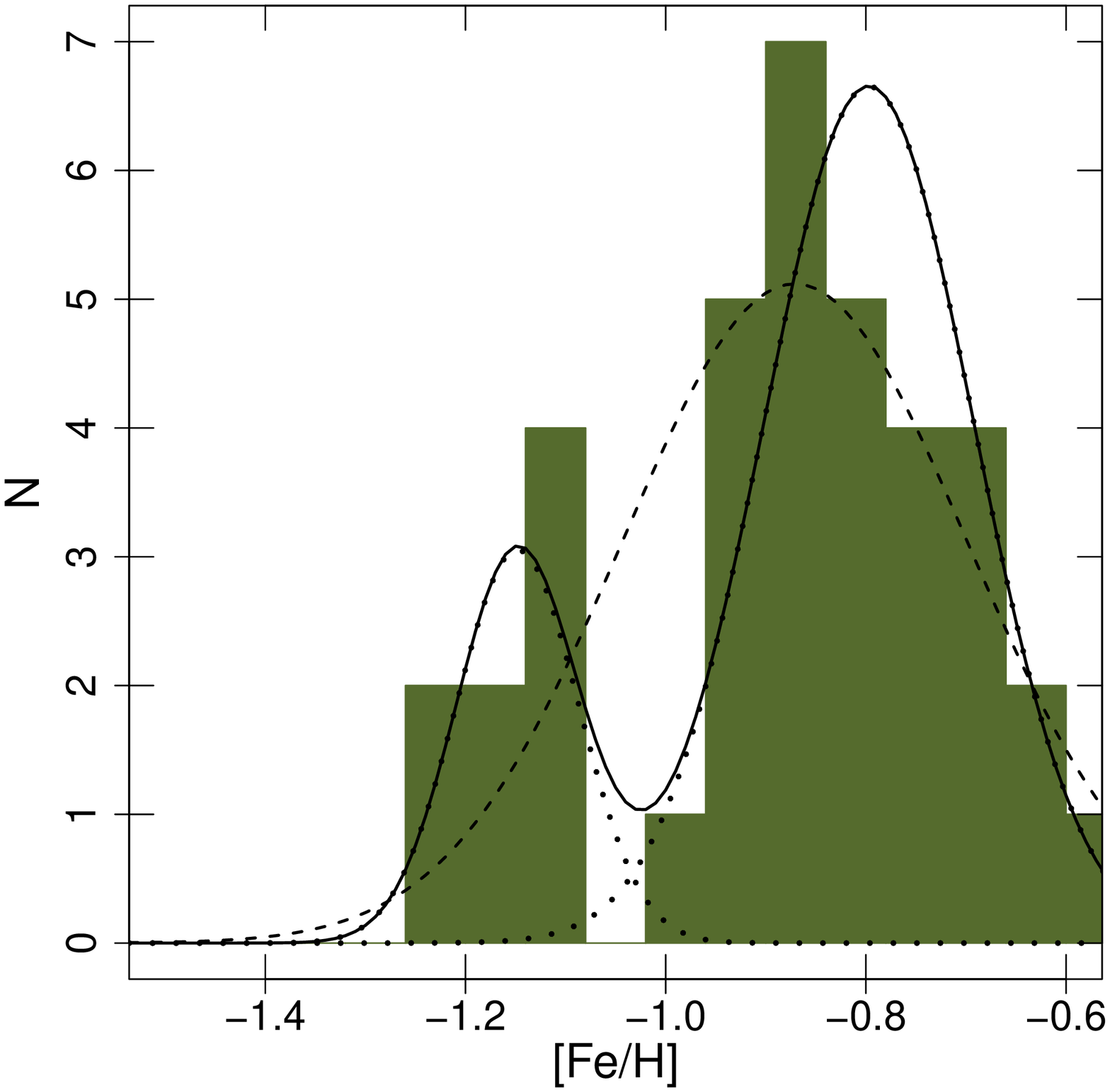}
\caption{Metallicity distribution of star clusters inside $a$=$3.4^{\circ}$. Solid, dotted and dashed lines show the same as in Figure\,\ref{fig.MD}.}
\label{fig.MDlt4}
\end{figure}

\begin{figure}
  \centering
\includegraphics[width=85mm]{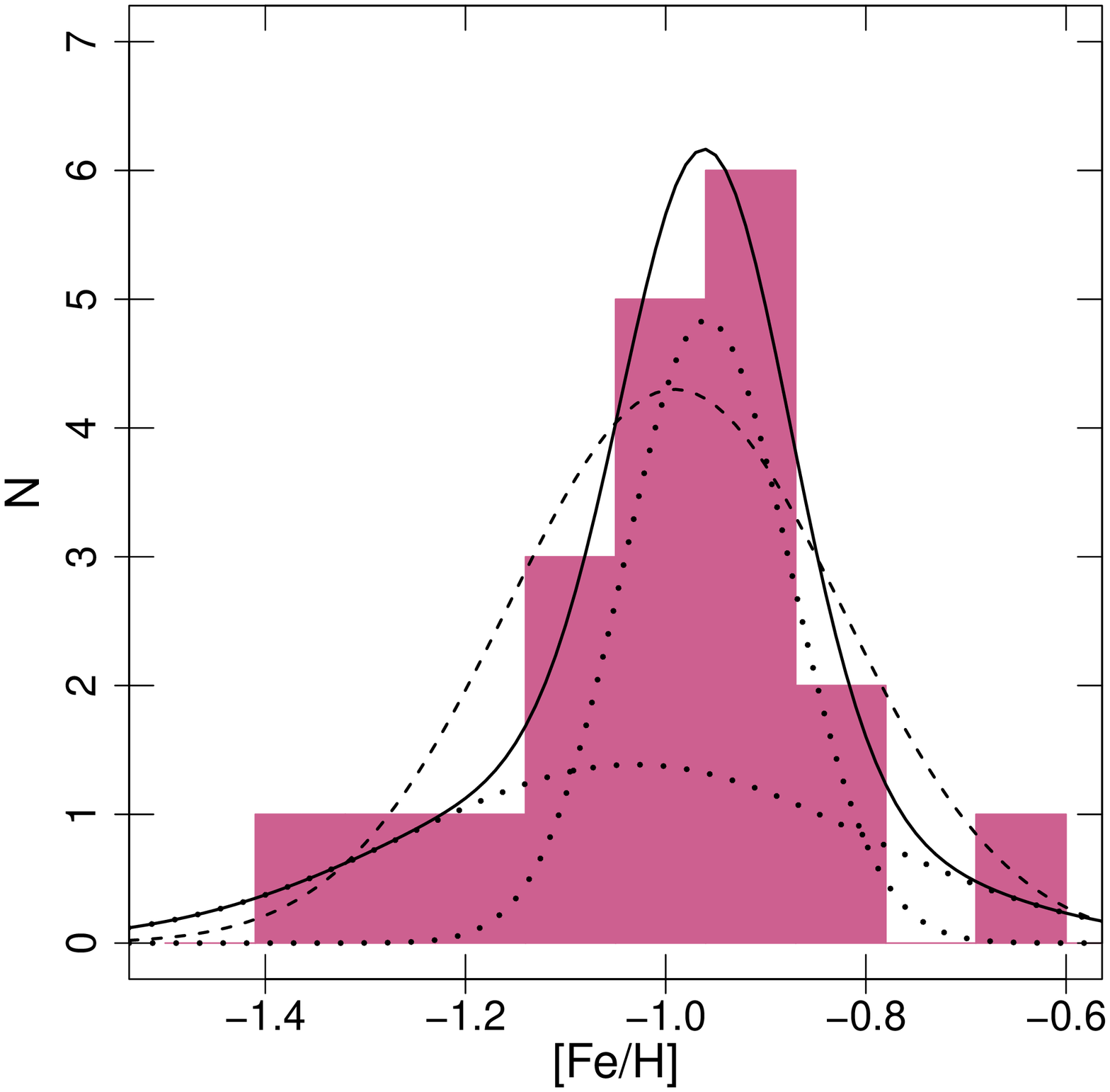}
\caption{Metallicity distribution of star clusters outside $a$=$3.4^{\circ}$. Solid, dotted and dashed lines show the same as in Figure\,\ref{fig.MD}.}
\label{fig.MDgt4}
\end{figure}

\begin{figure}
  \centering
\includegraphics[width=85mm]{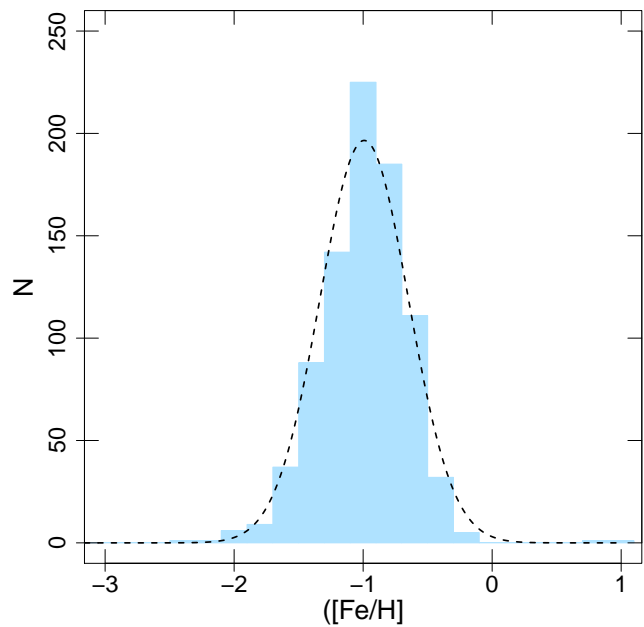}
\caption{Metallicity distribution of field stars. Dashed line shows an unimodal fit.}
\label{fig.MDfield}
\end{figure}

Considering the different behavior observed for the chemical properties of clusters and field stars in the inner and outer regions, we decided to analyse their MDs separately. We divided our total cluster sample considering clusters inside and outside of $3.4^{\circ}$ and applied the GMM test to each of these two subsamples (Table \ref{tab.gmm}). The internal and external MDs can be seen in Figures \ref{fig.MDlt4} and \ref{fig.MDgt4}, respectively.
As can be seen from the results included in the Table, while the outer MD presents a probability consistent with that found for the entire sample ($\sim$ 37\%), the inner part shows a high probability of a bimodal MD ($\sim$ 95\%). The inner MD analysis also shows a $DD$ greater than 2 and a negative kurtosis. This means that the possible bimodality in the internal region of the SMC is lost when we analyse the entire sample. The possible existence of bimodality in the inner region suggests the idea of two possible cluster populations coexisting towards the Main Body having a mean metallicity smaller and larger (-1.15 and -0.80 dex, respectively, see Table 4) than the typical mean metallicity of field stars ($\sim$ -1, P16). In order to analyse if the bimodality distribution is an artifact of our sample missing clusters with [Fe/H] $\sim$ -1 in the inner region, we performed an experiment generating random clusters with [Fe/H] values scattered around -1 dex and with a dispersion of 0.1, and repeated the GMM test. The results show that we had to add 14 clusters (which represents an increase of 37\% of the total sample) to bring the probability of bimodality down to 50\%. Even in these cases, the DD and the kurtosis remain larger than 2 and negative, respectively. So, we consider that the probability of bimodality being an artifact is low. In the case of the outer MD analysis shows a $DD$ lower than 2 and a positive kurtosis. The metallicities of our extended field star sample  clearly show a unimodal distribution (Figure \ref{fig.MDfield}), with a peak at -0.99, in agreement with previous works. Considering these results, we define two groups of clusters in the inner region, which we will call "metal-poor" and "metal-rich" clusters hereafter. The mean metallicities for the metal-rich and metal-poor clusters are -0.8 and -1.15 dex, respectively, corresponding to the peaks of the bimodal fit of the inner cluster MD (Table \ref{tab.gmm}).

\begin{table*}[!htb]
\caption{Gaussian Mixture Modeling results.}             
\label{tab.gmm}      
\centering                          
\footnotesize
\begin{tabular}{l c c c c c c c}        
\hline\hline                 
\noalign{\smallskip}
\multicolumn{1}{l}{Sample} &
\multicolumn{1}{c}{N} &
  \multicolumn{1}{c}{Unimodal fit} &
  \multicolumn{1}{c}{Bimodal fit} &
  \multicolumn{1}{c}{p value} &
  \multicolumn{1}{c}{Bimodality probability} &
\multicolumn{1}{c}{Peaks separation } &
\multicolumn{1}{c}{kurtosis } \\
\multicolumn{1}{c}{} &
    \multicolumn{1}{c}{} &
  \multicolumn{1}{c}{$\mu$($\sigma)$} &
  \multicolumn{1}{c}{ $\mu_1$($\sigma_1$); $\mu_2$ ($\sigma_2$)} &
  \multicolumn{1}{c}{} &
\multicolumn{1}{c}{parametric bootstrap } &
\multicolumn{1}{c}{} &
    \multicolumn{1}{c}{} \\
\noalign{\smallskip}
\hline\hline
\noalign{\smallskip}
All & 57 &-0.913 (0.180) & -0.830 (0.125); -1.106 (0.135) & 0.657 & 39.1\% &  2.83 $\pm$ 0.77 & -0.356 \\
Inner & 37 & -0.871 (0.173) & -0.797 (0.105); -1.148 (0.061) & 0.051 & 95.3\% & 4.17 $\pm$  0.40 & -0.632 \\
Outer  & 20 & -0.991 (0.167) & -0.959 (0.083); -1.031 (0.228) & 0.611 & 36.6\% & 1.91 $\pm$ 1.69 & 0.314 \\
\noalign{\smallskip}
\hline                                   
\end{tabular}\\
\end{table*}

\subsection{Metallicity Gradient}
\label{subsec.MG}

\begin{figure}
  \centering
\includegraphics[width=85mm]{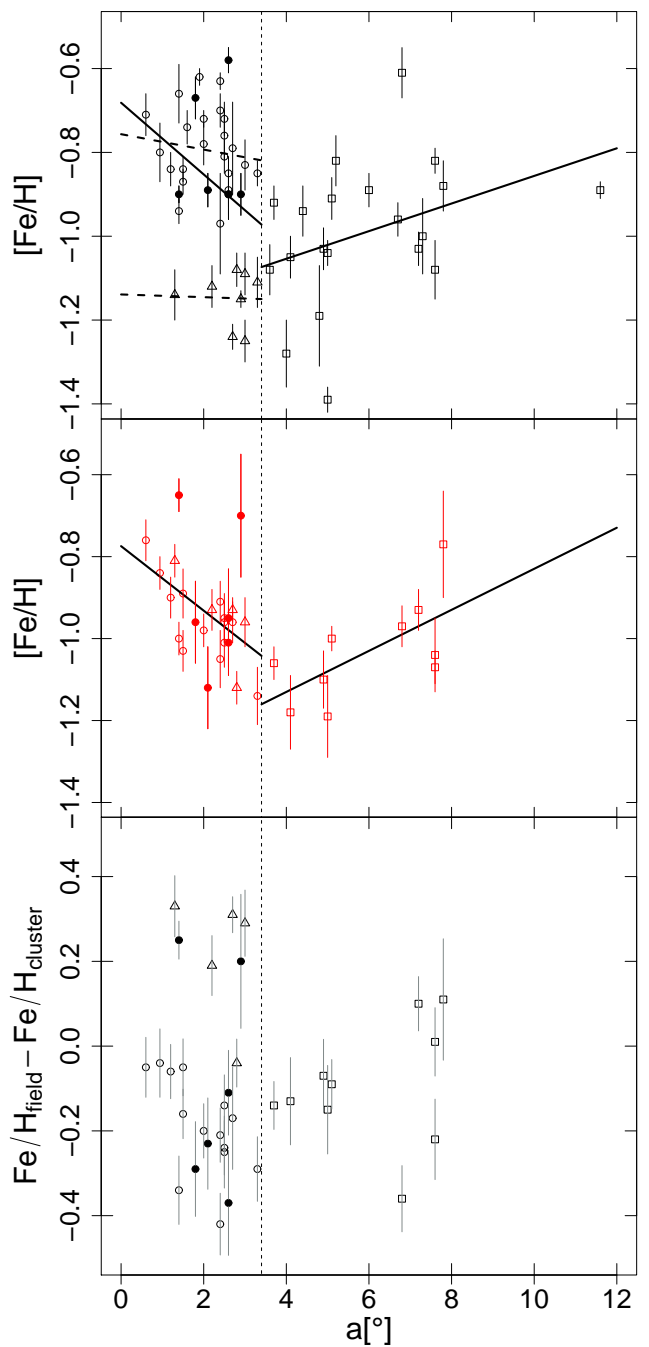} 
\caption{Metallicity gradient for clusters (upper panel) and field stars (middle panel). Bottom panel shows the difference between field and cluster metalicities. Filled and open circles are clusters and fields studied in this work and from literature, respectively. Circles and triangles show the metal-rich and metal-poor groups in the inner region, respectively. Squares show clusters in the outer region. The vertical line indicates the limit between the inner and outer regions.}
\label{fig.MG}
\end{figure}

The behaviour of the cluster metallicity as a function of the projected distance to the SMC centre $a$ for our cluster sample can be seen in the upper panel of Figure \ref{fig.MG}. The field MG is shown for comparison in the middle panel of that Figure while the lower panel shows the difference between field and cluster metallicities. This last panel has been included to analyse P10 and P16's suggestion that most clusters are more metal-rich than the corresponding fields, which can be observed for the inner clusters in the figure. This implies that the metallicity gradient of the clusters is shifted with respect to that of the field towards higher metallicities, with the exception of clusters of the metal-poor group. 

Following the analysis of our previous work, we determined from our extended samples the MGs in the inner and outer regions of the galaxy (solid lines in Fig. \ref{fig.MG}). We found  values of -0.08 $\pm$ 0.04 dex deg$^{-1}$ and 0.03 $\pm$ 0.02 dex deg$^{-1}$ for the inner and outer regions of the cluster sample, respectively. With respect to field stars, the linear fits  give a MG of -0.08 $\pm$ 0.03 dex deg$^{-1}$ for the inner part of the SMC and 0.05 $\pm$ 0.02 dex deg$^{-1}$ for the outer part. The inner field MG is in excellent agreement with the field MG found by other authors in that region also using CaT (-0.075 $\pm$ 0.011 dex deg$^{-1}$ from \citealt{dobbie+14b} and -0.08 $\pm$ 0.02 dex deg$^{-1}$ from P16). The inner cluster MG found in this work is consistent with P15 (-0.05 $\pm$ 0.04 dex deg$^{-1}$) considering errors, and it is in excellent agreement with P22 (-0.08 $\pm$ 0.04 dex deg$^{-1}$). The inner cluster MG is also consistent with the inner field MG. Although these results apparently lead to the conclusion that cluster and field stars have the same MG, the dispersion for field stars considering  median metallicities of field regions is very low in contrast to the high cluster metallicity dispersion, as can be clearly seen in Figure \ref{fig.MG}, particularly in the inner region. 

 P15 and P16 noted that two potential internal cluster groups are above and below the general field metallicity trend, which are now more clearly defined, as can be seen in Figure \ref{fig.MG}. These two potential groups correspond to the metal-poor and metal-rich cluster groups defined in the previous section. The mean metallicity and mean age values of the two groups, with their respective standard deviations, are -1.15 $\pm$ 0.06 dex and 4.2 $\pm$ 3.1 Gyr for the metal-poor group, and -0.8 $\pm$ 0.10 dex and 3.1 $\pm$ 1.7 Gyr for the metal-rich group.  
 
 We performed linear fits to the two internal groups separately (dashed lines in Figure \ref{fig.MG}, upper panel) yielding values for the MG of -0.02 $\pm$ 0.03 dex deg$^{-1}$ and -0.003 $\pm$ 0.042 dex deg$^{-1}$, for the metal-rich and metal-poor groups, respectively. Within the errors, the derived values are consistent with the absence of MG in the two potential internal groups. Therefore, the MG found for the inner clusters that is similar to that of the field stars, seems to be an artifact of the combination of two groups of clusters with a large spread in metallicity. 

It is noticeable in Figures \ref{fig.MDlt4} and \ref{fig.MG} that the inner SMC lacks clusters with [Fe/H] $\sim$ -1.0, whereas the outer clusters have metallicities around [Fe/H] $\sim$ -1.0. If we assume the existence of two groups of clusters with different origins in the inner region as real, then we can speculate in various ways. (1) Metal-rich clusters have formed in situ while the metal-poor clusters were accreted \citep{forbes+11}, as observed in elliptical galaxies \citep[e.g.][]{ennis+19,debortoli+20}, although the accretion mainly affects the outer part; (2)  Metal-poor and metal-rich gas formed clusters in the inner SMC whereas gas with intermediate metallicity formed clusters in the SMC outskirts; (3) The SMC experienced infalls of gas with different metallicities combined with inhomogeneous mixing of gas since its formation; the multiple encounters with the LMC would have triggered cluster formation by shock.

A completely different scenario to the existence of two inner populations is to postulate that the outer clusters were formed in the SMC during a continuous cluster formation with the interstellar medium chemical enrichment and inhomogeneous mixing, but the clusters with [Fe/H] $\sim$ -1.0 that would be concentrated in a given inner region were moved outwards by interactions with the LMC. In any case, all of these possible explanations are speculative. Real distances as well as dynamical simulations must be carried out to test these or other possible scenarios.

\subsection{Age-Metallicity Relation}

\begin{figure}
  \centering
\includegraphics[width=85mm]{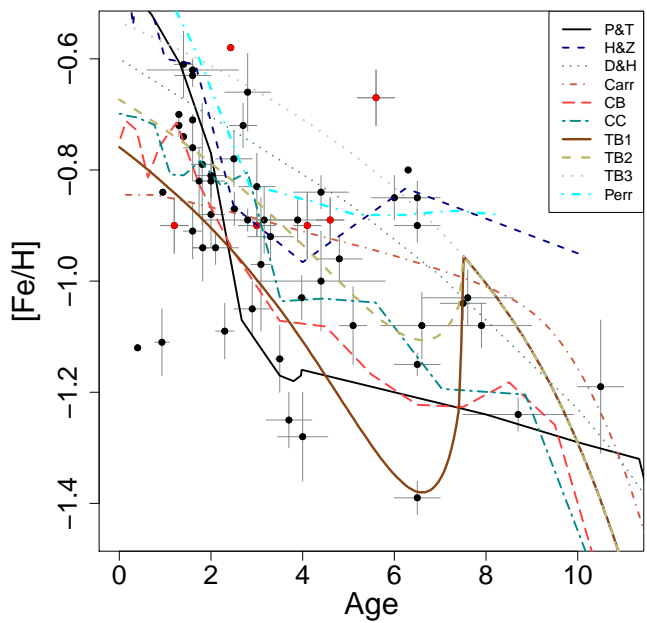}
\caption{Age-metallicity relation for the full cluster sample. Different AMR models are included. The clusters studied in this work and those taken from the literature are represented with red and black circles, respectively. 
References for the AMR models: P\&T: \citet{pagel+98}, H\&Z: \citet{harris+04}, D\&H: \citet{dacosta+98}, Carr:\citet{carrera+08}, CB and CC: \citet{cignoni+13}, T\&B1, T\&B2 and T\&B3: \citet{tsujimoto+09}, Perr: \citet{perren+17}.}
\label{fig.AMR_all}
\end{figure}

\begin{figure}
  \centering
\includegraphics[width=85mm]{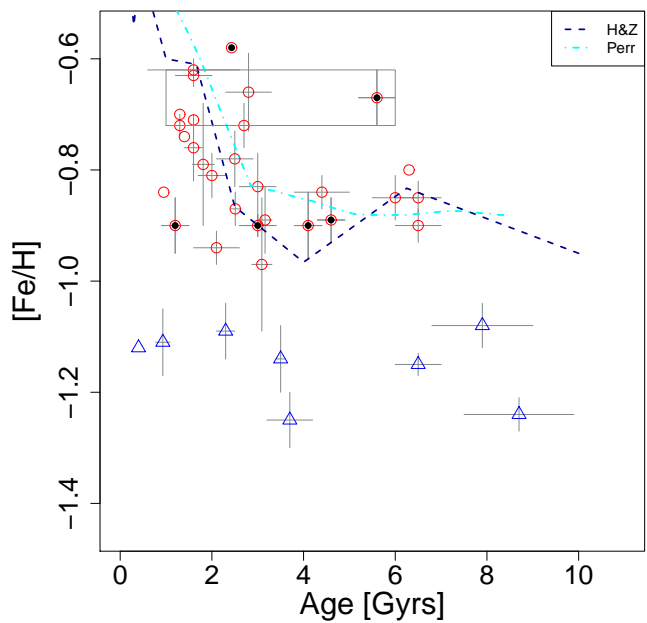}
\caption{Age-metallicity relation inside $3.4^{\circ}$. Red circles and blue triangles show those clusters corresponding to the metal-rich and metal-poor groups, respectively. The six clusters studied in this work are represented with filled symbols. The references for the AMR models are the same as in Figure \ref{fig.AMR_all}. The cluster HW41 has been graphed within a rectangle that represents the range of ages that can be found in the literature for this cluster (see the text for details.)}
\label{fig.AMR_ltrt}
\end{figure}

In an effort to help constrain the chemical evolution of the SMC, we plot in Figure \ref{fig.AMR_all} the AMR for the full cluster sample which shows, as in our previous work, that that none of available models of chemical evolution reproduce the data adequately. If the different regions of the SMC (D16,D21) are analysed separately, the AMR becomes somewhat clearer.
The AMR of the Main Body, which is the region of interest of the present paper, is presented in Figure \ref{fig.AMR_ltrt}. In that figure we have distinguished with red and blue symbols clusters belonging to the metal-rich and metal-poor groups, respectively. We compare the observational data with different models of chemical evolution (see the caption of the figure for model references).  

It is noticeable in Figure \ref{fig.AMR_ltrt} that 21 of the 28 metal-rich clusters (75\%) are concentrated at ages younger than 4-5 Gyr and none are older than about 6-7 Gyr, whereas metal-poor clusters cover a wider age range, although the number of metal-poor clusters is low. The age of $\sim$ 6 Gyr corresponds approximately to the time where star formation was probably triggered by the infall to the SMC-LMC pair towards the Milky Way \citep[e.g.][]{besla+12}. It is interesting to note that the age distribution of the SMC clusters shows a peak at $\sim$ 5 Gyrs \citep{piatti+11,parisi+14,bica+20} which agrees with a peak in the fields SFR of $log(t)$ = 9.7 from \citet{rubele+18}.  On the other hand, while the clusters of the metal-rich group  present a metallicity dispersion of $\sim$ 0.4 dex, the clusters of the metal-poor group are found concentrated in a significantly smaller metallicity range ($\pm$ 0.1 dex) around the previously calculated mean value of -1.15 dex. The oldest clusters of the metal-rich group show a primordial constant metallicity of -0.86 $\pm$ 0.04 dex (without considering HW41, see text below) until 4-5 Gyr ago, showing a considerable subsequent chemical enrichment process which increased the metallicity considerably, at least for about half of the younger clusters. On the contrary, metal-poor cluster formation appears to not have undergone any chemical enrichment throughout the life of the SMC, which suggests a low efficient gas mixing for the metal-poor gas that kept forming clusters during the entire life of the SMC. The chemical evolution of the metal-rich group clusters would seem to be well represented by the \citet[][H\&Z]{harris+04} and \citet[][]{perren+17} models. Particularly interesting is the fact that the AMR proposed by H\&Z was specifically derived from the Star Formation History (SFH) of 361 regions located in the main body of the SMC (4$^{\circ} \times$ 4.5$^{\circ}$). This finding supports the arguments of D16  that each region of the SMC should be analysed separately, possibly revealing specific pieces of the SMC chemical evolution history. This result also suggests a similar chemical evolution for field stars analysed by H\&Z and and metal-rich star clusters analysed here in the SMC main body region.

One exception to this apparent match with the H\&Z and \citet{perren+17} models is the cluster HW\,41 for which there is a large dispersion in the age determination that can be found in the literature (marked in Figure \ref{fig.AMR_ltrt} with a rectangle). \citet{piatti11a} derived an age of 6 $\pm$ 1 Gyr using the $\delta$T$_1$ age index in the Washington photometric system and \citet{parisi+14} determined an age of 5.6 $\pm$ 0.4 from the conversion of Piatti's $\delta$T$_1$ to $\delta$V index. On the other hand, \citet{glatt+10} and \citet{perren+17}, applying automatic fitting methods, found age values for HW\,41 of 1 Gyr and 3.8 Gyr, respectively. We adopted for our analysis the age from \citet{parisi+14} but a more reliable age needs to be derived for this cluster.  

Considering the analysis in this section and, if we assume that the metal-rich group corresponds to clusters formed in situ while those of the metal-poor group have a different origin, then the AMR of the SMC Main Body corresponds to the relationship found for the metal-rich group.


\section{SUMMARY AND CONCLUSIONS}
\label{sec.conclu}

Using Ca II Triplet lines we have determined the mean RV and metallicity of 6 clusters and 6 fields in the SMC. We added another 51 clusters and  30 fields studied with the same technique and with metallicities on the same scale. We divided our samples in inner and outer regions considering the breakpoint derived by \citet{dias+21} ($a = 3.4^{\circ}$). According to the definition of the \citet[]{dias+16b,dias+21} and P22, the inner region corresponds to the SMC Main Body. Our main conclusions are the following:

\begin{itemize}

\item We find a high probability (95.3\%) that the cluster MD is bimodal in the inner region but unimodal in the outer regions. 
\item Considering the bimodal MD in  the inner region we define two cluster groups as metal-rich and metal-poor groups with mean metallicities of -0.80 and -1.15 dex, respectively.
\item Outer cluster MD and field MD have coincident, unimodal peaks at -0.99 dex.
\item Cluster metallicity gradient (MG) is negative (-0.08$\pm$0.04 dex deg$^{-1}$) in the inner region but positive or null in the outer region (0.03$\pm$0.02 dex deg$^{-1}$), in agreement with the MG for field stars. However, linear fits for the metal-rich and metal-poor clusters separately are consistent with no MG. In the outer region field stars MG is significantly possitive.
\item With our extended sample we continue to find the cluster metallicity gap in the Main Body at $\sim$ -1, as suggested by P16.
\item The Age-Metallicity Relation (AMR) in the inner region shows that the metal-rich clusters appear to follow the chemical enrichment of field stars by \citet{harris+04} or the model proposed by \citet{perren+17}, but metal-poor clusters do not present any chemical enrichment.

\end{itemize}

In this work we present observational evidence that the chemical enrichment is complex in the SMC Main Body. Two cluster groups with potential different origins could be coexisting in the Main  Body but more data with precise and homogeneous metallicities and distances are needed to corroborate not only the metallicity gap but also any possible projection effect. Dynamical simulations are required to understand possible different origins for the metal-rich and metal-poor cluster groups in the SMC Main Body.

\begin{acknowledgements}

We thank the comments of the referee, which helped to improve this paper.This work was funded with grants from Consejo Nacional de Investigaciones Cient\'{\i}ficas y T\'ecnicas de la Rep\'ublica Argentina, Agencia Nacional de   Promoci\'on Cient\'{\i}fica y Tecnol\'ogica, and Universidad Nacional de La Plata (Argentina). This research was partially supported by the Argentinian institution SECYT (Universidad Nacional de Córdoba).
Based on observations obtained at the Gemini Observatory (programme  GS-2016B-Q-17, PI: M.C. Parisi), which is operated by the Association of Universities for Research in Astronomy, Inc., under a cooperative agreement with the NSF on behalf of the Gemini partnership: the National Science Foundation (United States), the National Research Council (Canada), CONICYT (Chile), the Australian Research Council (Australia), Ministério da Ciencia, Tecnologia e Inovacao (Brazil) and Ministerio de Ciencia, Tecnología e Innovación Productiva (Argentina). This work presents results from the European Space Agency (ESA) space mission Gaia. Gaia data are being processed by the Gaia Data Processing and Analysis Consortium (DPAC). Funding for the DPAC is provided by national institutions, in particular the institutions participating in the Gaia MultiLateral Agreement (MLA). The Gaia mission website is https://www.cosmos.esa.int/gaia. The Gaia archive website is https://archives.esac.esa.int/gaia. This research has made use of "Aladin sky atlas" developed at CDS, Strasbourg Observatory, France. D.G. gratefully acknowledges support from the ANID BASAL project ACE210002.
D.G. also acknowledges financial support from the Direcci\'on de Investigaci\'on y Desarrollo de
la Universidad de La Serena through the Programa de Incentivo a la Investigaci\'on de
Acad\'emicos (PIA-DIDULS). B.D. acknowledges support by FONDECYT iniciación grant No. 11221366.

\end{acknowledgements}

   \bibliographystyle{aa} 
   \bibliography{bibliography} 

\appendix

\end{document}